\documentclass{PoS}

\usepackage{amsmath}
\allowdisplaybreaks

\let\OLDthebibliography\thebibliography
\renewcommand\thebibliography[1]{
  \OLDthebibliography{#1}
  \setlength{\parskip}{0pt}
  \setlength{\itemsep}{2pt plus 0.3ex}
}

\title{Hadronic Higgs boson decay at order $\alpha_s^4$ and $\alpha_s^5$}

\ShortTitle{Hadronic Higgs boson decay at order $\alpha_s^4$ and $\alpha_s^5$}

\author{\speaker{Joshua Davies}\\
        Institut f\"ur Theoretische Teilchenphysik, Karlsruhe Institute of Technology (KIT)\\
        76128 Karlsruhe, Germany\\
        E-mail: \email{joshua.davies@kit.edu}}
\author{Matthias Steinhauser\\
        Institut f\"ur Theoretische Teilchenphysik, Karlsruhe Institute of Technology (KIT)\\
        76128 Karlsruhe, Germany\\
        E-mail: \email{matthias.steinhauser@kit.edu}}
\author{David Wellmann\\
        Institut f\"ur Theoretische Teilchenphysik, Karlsruhe Institute of Technology (KIT)\\
        76128 Karlsruhe, Germany\\
        E-mail: \email{david.wellmann@kit.edu}}

\abstract{
	We compute corrections to the decay of the Standard Model Higgs boson to hadrons, to the fourth order in the strong coupling constant $\als$. We use an effective theory in which the Higgs boson couples directly to bottom quarks and to gluons, via top quark--mediated effective couplings. Numerically, our results are of a comparable size to the previously-known ``massless'' contributions and complete the order $\als^4$ corrections to the hadronic decay of the Higgs boson. In these proceedings we also provide an independent cross check of the gluonic Higgs boson decay at order $\als^5$.
}

\FullConference{XXV International Workshop on Deep-Inelastic Scattering and Related Subjects\\
		3-7 April 2017\\
		University of Birmingham, UK}

\newcommand{\als}{\alpha_s}
\newcommand{\alsfivepi}{a_s}

\newcommand{\Deloo}{\Delta_{11}}
\newcommand{\Delot}{\Delta_{12}}
\newcommand{\Deltt}{\Delta_{22}}
\newcommand{\Dlight}{\Delta_{\rm light}}
\newcommand{\Dtop}{\Delta_{\rm top}}
\newcommand{\Dggmbz}{\Delta_{gg}^{\mb=0}}

\newcommand{\Pioo}{\Pi_{11}}
\newcommand{\Piot}{\Pi_{12}}
\newcommand{\Pito}{\Pi_{21}}
\newcommand{\Pitt}{\Pi_{22}}

\newcommand{\Oonep}{{\cal O}_1^\prime}
\newcommand{\Otwop}{{\cal O}_2^\prime}
\newcommand{\Cone}{C_1}
\newcommand{\Ctwo}{C_2}

\newcommand{\LH}{L_H}
\newcommand{\LHtwo}{L^{\;2}_H}
\newcommand{\LHthr}{L^{\;3}_H}

\newcommand{\MH}{M_H}
\newcommand{\MHtwo}{M^{\;2}_H}
\newcommand{\MHthr}{M^{\;3}_H}
\newcommand{\MHfou}{M^{\;4}_H}

\newcommand{\mb}{m_b}
\newcommand{\mbtwo}{m^{\;2}_b}

\newcommand{\Mt}{M_t}
\newcommand{\Mttwo}{M^{\;2}_t}

\newcommand{\cf}{C_F}
\newcommand{\cftwo}{C^{\,2}_F}

\newcommand{\ca}{C_A}
\newcommand{\catwo}{C^{\,2}_A}
\newcommand{\cathr}{C^{\,3}_A}
\newcommand{\nl}{n_l}
\newcommand{\nltwo}{n^{\,2}_l}
\newcommand{\nlthr}{n^{\,3}_l}

\newcommand{\zetatwo}{\zeta_2}
\newcommand{\zetathr}{\zeta_3}

\newcommand{\zetafiv}{\zeta_5}

\newcommand{\frct}[2]{\mbox{\Large{$\frac{#1}{#2}$}}}

\begin{document}

\section{Introduction}
	In the coming years, it will be an important task to precisely determine the Higgs boson couplings to other Standard Model particles. To this end, it is important to have a good theoretical understanding of the decay rate of the Higgs boson into bottom quarks. Combined with the Higgs boson's decay rate into gluons, it forms some 70\% of the hadronic decay width and therefore influences all Higgs boson branching ratios.

	QCD corrections to the partial width $\Gamma(H\to b\bar{b})$ have been studied for some time. At one and two loops they are known with exact bottom quark--mass dependence~\cite{Braaten:1980yq,Gorishnii:1990zu,Chetyrkin:1997mb,Harlander:1997xa,Chetyrkin:1998ix}. Beyond this order, at three and four loops, only the ``massless'' approximation was known~\cite{Chetyrkin:1996sr,Chetyrkin:1997vj,Baikov:2005rw}. In Ref.~\cite{Davies:2017xsp} the order $\als^4$ corrections were completed. The calculation will be described in these proceedings. For a summary of further corrections, the reader is referred to~\cite{deFlorian:2016spz,Spira:2016ztx,Djouadi:1997yw}.

	For the calculation in the limit $\Mt\gg\MH$, we consider an effective Lagrange density in which the top quark has been integrated out. It consists of QCD and effective couplings between the Higgs boson and bottom quarks and gluons,
	\begin{equation}
		{\cal L}_{\rm eff} = -\frac{H^0}{v^0} \left( \Cone [\Oonep] + \Ctwo [\Otwop] \right) + {\cal L}_{\rm QCD}^\prime,
	\label{eq:efflag}
	\end{equation}
	where primed quantities refer to the five-flavour effective theory. The bare Higgs field $H^0$ and vacuum expectation $v^0$ may be identified with their renormalized quantities since we neglect electroweak effects in this calculation. The top quark--mass dependence is provided by the coefficient functions $\Cone$ and $\Ctwo$, which are known to fifth order in $\als$~\cite{Schroder:2005hy,Chetyrkin:2005ia,Liu:2015fxa}.
	The bare effective operators are given by $\Oonep = (G_{a,\mu\nu}^{0\prime})^2$ and $\Otwop = m_b^{0\prime}\bar{b}^{0\prime}b^{0\prime}$, and $[\Oonep]$, $[\Otwop]$ are their renormalized counterparts.

	Correlators formed by the above operators can be related, via the optical theorem, to the total decay rate of the Higgs boson. With the correlator
	\begin{equation}
		\Pi_{ij}(q^2) = i\int {\rm d} x e^{iqx} \langle 0 | T[{\cal O}_i^\prime,{\cal O}_j^\prime ] |0 \rangle,
	\label{eq:definecorrelator}
	\end{equation}
	we define the quantities
	\begin{align}
		&\Deloo = \mbox{Im}\left[\Pioo(\MHtwo)\right]/(32\pi\MHfou),\qquad
		\Delot = \mbox{Im}\left[\Piot(\MHtwo)+\Pito(\MHtwo)\right]/(6\pi\MHtwo\mbtwo),\nonumber\\
		&\Deltt = \mbox{Im}\left[\Pitt(\MHtwo)\right]/(6\pi\MHtwo\mbtwo),
	\label{eq:defdelta}
	\end{align}
	in terms of which,
	\begin{equation}
		\Gamma(H\to\mbox{hadrons}) = A_{b\bar{b}} \left[\left(C_2\right)^2 \left(1+\Deltt\right)
		                             + C_1 C_2 \Delot \right]
                                     + A_{gg} \left(C_1\right)^2 \Deloo
	\label{eq:decaywidth}
	\end{equation}
	where $A_{b\bar{b}} = 3 G_F \MH \mbtwo(\mu)/(4\pi\sqrt{2})$ and $A_{gg} = 4 G_F \MHthr/(\pi\sqrt{2})$.

	$\Pioo$ and $\Pitt$ are already known at four~\cite{Baikov:2006ch} and five~\cite{Baikov:2005rw} loops, providing $\als^5$ and $\als^4$ corrections to the decay rate, respectively. The mixed contribution $\Piot(=\Pito)$, however, was known only at three loops~\cite{Chetyrkin:1997vj}. It has now been computed at four loops~\cite{Davies:2017xsp}, completing the $\als^4$ QCD corrections to the decay rate. Additionally, we compute new two-, three- and four-loop corrections to $\Pioo$ which are suppressed by one power of $(\mbtwo/\MHtwo)$; they are numerically small, but are of the same order in $(\mbtwo/\MHtwo)$ as the leading contributions from $\Piot$ and $\Pitt$. The results presented in these proceedings (Eq.~(\ref{eq:Deloo4})) provide an independent cross check of the four-loop $(\mbtwo/\MHtwo)^0$ terms of $\Pioo$.

	Corrections suppressed by powers of $(\MHtwo/\Mttwo)$ are small, and not considered here.


\section{Calculation}
	The calculation was performed by making use of a well-tested collection of software. We use \texttt{qgraf}~\cite{Nogueira:1991ex} to generate the diagrams. \texttt{q2e} and \texttt{exp}~\cite{Harlander:1997zb,Seidensticker:1999bb,q2eexp} are used to map the diagrams into a set of two-point integral topologies. \texttt{FORM}~\cite{Kuipers:2012rf} is used to perform the Dirac traces and to re-write all integrals as linear combinations of scalar integrals.
	The resulting scalar integrals are reduced to master integrals by \texttt{FIRE 5.2}~\cite{Smirnov:2014hma,FIRE}; we find 28 four-loop master integrals, which have been computed in Refs.~\cite{Baikov:2010hf,Smirnov:2010hd,Lee:2011jt}. We do not discuss the details of the renormalization of the bare expressions here, and refer the reader to Ref.~\cite{Davies:2017xsp}.

	Using this setup, we have computed one-, two-, three- and four-loop corrections to each of the correlators $\Pioo$, $\Piot$ and $\Pitt$. In the case of $\Pioo$ we compute both the leading $(\mbtwo/\MHtwo)^0$ and sub-leading $(\mbtwo/\MHtwo)$ contributions. In all cases we employ a generic gauge parameter $\xi$ and expand the amplitudes to linear order in $\xi$ before integral reduction. All dependence on $\xi$ drops out after reduction to a minimal set of master integrals.


\section{Results}
	To investigate the numerical size of the new corrections, we re-write Eq.~(\ref{eq:decaywidth}) in the form
	\begin{equation}
		\Gamma(H\to\mbox{hadrons}) = A_{b\bar{b}}\left(1 + \Dlight + \Dtop + \Dggmbz\right),
	\end{equation}
	with a common prefactor. $\Dlight$ gives the terms for $\Cone = 0$ and $\Ctwo = 1$, i.e., terms with no top quark--mass dependence. $\Dggmbz$ collects the leading terms from $\Deloo$, which are proportional to $(\mbtwo/\MHtwo)^0$. The remaining terms are collected inside $\Dtop$. For completeness, these are defined in terms of the symbols of Eq.~(\ref{eq:decaywidth}) as
	\begin{align}
		&\Dlight = \Deltt, \qquad\qquad \Dggmbz = \frac{16\MHtwo}{3\mbtwo}(\Cone)^2 \Deloo^{\mb=0},\nonumber\\
		&\Dtop = \left[(\Ctwo)^2-1\right](1+\Deltt) + \Cone\Ctwo\Delot + \frac{16\MHtwo}{3\mbtwo}(\Cone)^2 \Deloo^{\mbtwo}.
	\end{align}
	Numerically, we find
	\begin{align}
		\Dlight \approx & \:
			5.6667 \alsfivepi  + 29.1467 \alsfivepi^2 + 41.7576 \alsfivepi^3 - 825.7466 \alsfivepi^4
			\nonumber\\*
			\approx & \: 0.2033 + 0.03752 + 0.001929 - 0.001368,
	\label{eq::Delta_light}\\[3mm]
		\Dtop \approx & \:
			\alsfivepi^2 \, [2.5556_{12} + 0.9895_{22} ]
\nonumber\\*&\mbox{}
			+ \alsfivepi^3 \, [0.2222_{11} + 42.1626_{12} + 13.0855_{22}]
\nonumber\\*&\mbox{}
			  + \alsfivepi^4 \, [8.3399_{11} + 338.9021_{12} + 50.6346_{22}]
\nonumber\\*&\mbox{}
              + \alsfivepi^5 \, [137.145_{11}]
\nonumber\\*
		\approx & \:
			0.003290_{12} + 0.001274_{22} 
\nonumber\\*&\mbox{}
			+ 0.00001026_{11} + 0.001947_{12} + 0.0006043_{22}
\nonumber\\*&\mbox{}
			+ 0.00001382_{11} + 0.0005616_{12} + 0.00008390_{22}
\nonumber\\*&\mbox{}
            + 0.00000815_{11},
	\label{eq::Delta_top}
\\[3mm]
		\Dggmbz \approx & \:
			{ \frac{\MHtwo}{27\mbtwo} \left(
				\alsfivepi^2 + 17.9167 \alsfivepi^3 + 153.0921 \alsfivepi^4
				+ 392.6176 \alsfivepi^5
			\right)}
\nonumber\\*
		\approx & \:
			0.09699 + 0.06235 + 0.01911 + 0.001759,
	\label{eq::Delta_gg}
	\end{align}
	in which the expansion parameter is $\alsfivepi = \alpha_s^{(5)}(\mu)/\pi$ and we have used the values $\alpha_s^{(5)}(\MH) = 0.1127$, $\mb(\MH) = 2.773$ GeV, $\MH = 125.09$ GeV, $\Mt = 173.21$ GeV. The subscripts $_{11}$, $_{12}$ and $_{22}$ in Eq.~(\ref{eq::Delta_top}) tag the origin of each number. $\Dggmbz$ includes the $\alsfivepi^5$ term from Ref.~\cite{Baikov:2006ch} and $\Dtop$ the new $\alsfivepi^5(\mbtwo/\MHtwo)$ term computed here.

	We observe that the $\alsfivepi^4$ term of $\Dlight$ is only about 30\% smaller than the $\alsfivepi^3$ term. For this reason it is important to consider also the complete $\alsfivepi^4$ term of $\Dtop$, as we have done here. It is of the same order of magnitude, but appears with an opposite sign. Its inclusion significantly improves the convergence of the sum of all $(\mbtwo/\MHtwo)$ terms,
	\begin{equation}
		1+\Dlight+\Dtop \approx 1 + 0.2033 + 0.04208 + 0.004490 - 0.0007090,
	\end{equation}
	to which the $\alsfivepi^4$ term now provides a correction of just 0.0567\%. The $(\mbtwo/\MHtwo)$ corrections to $\Dtop$ are very small.

	We now present the $\alsfivepi^3$ corrections to $\Deloo$, which have not been fully presented before, in terms of their colour factors and including the scale-dependence. The $(\mbtwo/\MHtwo)$ terms are new. We find
	\begin{align}
\Deloo\bigg|_{\alsfivepi^3} = &\:\:
        \cathr \* \bigg(
            \frct{15420961}{46656}
          - \frct{704}{9}\*\zetatwo
          - \frct{44539}{432}\*\zetathr
          + \frct{385}{24}\*\zetafiv
          + \Big[\frct{965285}{5184} - \frct{1331}{72}\*\zetatwo - \frct{605}{24}\*\zetathr\Big]\*\LH
\nonumber\\[-2mm]&
          + \frct{352}{9}\*\LHtwo
          + \frct{1331}{432}\*\LHthr
          \bigg)
       - \nl\*\catwo \* \bigg(
            \frct{667627}{3888}
          - \frct{2021}{48}\*\zetatwo
          - \frct{2443}{144}\*\zetathr
          + \frct{5}{12}\*\zetafiv
          + \Big[\frct{170263}{1728}
\nonumber\\[-2mm]&
                 - \frct{121}{12}\*\zetatwo - \frct{11}{3}\*\zetathr\Big]\*\LH
          + \frct{2021}{96}\*\LHtwo
          + \frct{121}{72}\*\LHthr
          \bigg)
       - \nl\*\cf\*\ca \* \bigg(
            \frct{23221}{576}
          - \frct{143}{48}\*\zetatwo
          - \frct{341}{16}\*\zetathr
          - \frct{5}{2}\*\zetafiv
\nonumber\\[-2mm]&
          + \Big[\frct{8569}{576} - \frct{11}{2}\*\zetathr\Big]\*\LH
          + \frct{143}{96}\*\LHtwo
          \bigg)
       + \nl\*\cftwo \* \bigg(
            \frct{221}{192}
          + 3\*\zetathr
          - 5\*\zetafiv
          + \frct{3}{32}\*\LH
          \bigg)
       + \nltwo\*\ca \* \bigg(
            \frct{103327}{3888}
\nonumber\\[-2mm]&
          - \frct{173}{24}\*\zetatwo
          + \frct{7}{72}\*\zetathr
          + \Big[\frct{9187}{576} - \frct{11}{6}\*\zetatwo + \frct{1}{6}\*\zetathr\Big]\*\LH
          + \frct{173}{48}\*\LHtwo
          + \frct{11}{36}\*\LHthr
          \bigg)
       + \nltwo\*\cf \* \bigg(
            \frct{55}{8}
          - \frct{13}{24}\*\zetatwo
\nonumber\\[-2mm]&
          - \frct{15}{4}\*\zetathr
          + \Big[\frct{767}{288} - \zetathr\Big]\*\LH
          + \frct{13}{48}\*\LHtwo
          \bigg)
       - \nlthr \* \bigg(
            \frct{7127}{5832}
          - \frct{7}{18}\*\zetatwo
          - \frct{1}{27}\*\zetathr
          + \Big[\frct{127}{162} - \frct{1}{9}\*\zetatwo\Big]\*\LH
\nonumber\\[-2mm]&
          + \frct{7}{36}\*\LHtwo
          + \frct{1}{54}\*\LHthr
          \bigg)
\nonumber\\
+ \frct{\mbtwo}{\MHtwo} & \* \bigg\{
         \catwo \* \bigg(
            \frct{991}{2}
          - \frct{121}{2}\*\zetatwo
          - \frct{317}{24}\*\zetathr
          - \frct{35}{8}\*\zetafiv
          + \Big[\frct{2675}{12} + 22\*\zetathr\Big]\*\LH
          + \frct{121}{4}\*\LHtwo
          \bigg)
       + \cf\*\ca \* \bigg(
            \frct{148921}{192}
\nonumber\\[-2mm]&
          - \frct{1415}{8}\*\zetatwo
          - \frct{3881}{24}\*\zetathr
          - \frct{685}{24}\*\zetafiv
          + \Big[\frct{42767}{96} - \frct{99}{4}\*\zetatwo - \frct{189}{4}\*\zetathr\Big]\*\LH
          + \frct{1415}{16}\*\LHtwo
          + \frct{33}{8}\*\LHthr
          \bigg)
\nonumber\\[-2mm]&
       + \cftwo \* \bigg(
            \frct{18485}{64}
          - \frct{693}{8}\*\zetatwo
          - \frct{675}{8}\*\zetathr
          + \frct{105}{4}\*\zetafiv
          + \Big[\frct{5319}{32} - \frct{81}{4}\*\zetatwo - 27\*\zetathr\Big]\*\LH
          + \frct{693}{16}\*\LHtwo
\nonumber\\[-2mm]&
          + \frct{27}{8}\*\LHthr
          \bigg)
       - \nl\*\ca \* \bigg(
            \frct{20135}{144}
          - 22\*\zetatwo
          + \frct{367}{24}\*\zetathr
          + \Big[\frct{845}{12} + 4\*\zetathr\Big]\*\LH
          + 11\*\LHtwo
          \bigg)
       - \nl\*\cf \* \bigg(
            \frct{13991}{96}
\nonumber\\[-2mm]&
          - \frct{121}{4}\*\zetatwo
          - \frct{133}{3}\*\zetathr
          + \frct{10}{3}\*\zetafiv
          + \Big[\frct{3475}{48} - \frct{9}{2}\*\zetatwo - 9\*\zetathr\Big]\*\LH
          + \frct{121}{8}\*\LHtwo
          + \frct{3}{4}\*\LHthr
          \bigg)
       + \nltwo \* \bigg(
            \frct{157}{18}
          - 2\*\zetatwo
\nonumber\\[-2mm]&
          + 5\*\LH
          + \LHtwo
          \bigg)
\bigg\}.
	\label{eq:Deloo4}
	\end{align}
	Here, $\ca$ and $\cf$ are the quadratic Casimir invariants of $SU(N)$, $\LH = \log(\mu^2/\MHtwo)$, $\nl$ is the number of light quarks running in loops and $\zeta_n$ is the Riemann Zeta function.
	Eq.~(\ref{eq:Deloo4}) agrees with the expression presented in~\cite{Baikov:2006ch} if one sets $\LH = 0$, $\ca = 3$, $\cf = 4/3$ and $(\mbtwo/\MHtwo) = 0$. We also find agreement with~\cite{Moch:2007tx} if one sets $\LH = 0$ and $(\mbtwo/\MHtwo) = 0$, which contains an independent calculation of all but the $\zetatwo$ terms of $\Deloo|_{\alsfivepi^3}$.

\section{Conclusions}
	By computing the previously-unknown $\als^4$ corrections to $\Delot$, we complete the QCD corrections to the hadronic decay of the Higgs boson at this order. The computation is performed in an effective theory, however additional corrections proportional to powers of $(\MHtwo/\Mttwo)$ are known to be small. The new contributions to $\Gamma(H\to\mbox{hadrons})$ are of the same order of magnitude as the previously-known contributions, but come with an opposite sign; they do much to improve the convergence of the QCD perturbative series.

	Additionally, we provide an independent cross check of the four-loop Higgs boson decay to gluons. Computer-readable files containing the analytic results of these proceedings and Ref.~\cite{Davies:2017xsp} can be found at~\cite{progdata}.

\section*{Acknowledgements}
This work is supported by the BMBF grant 05H15VKCCA.

\end{document}